\documentclass[manuscript]{aastex6}
\usepackage{amsmath}
\newcommand{\refeq}[1]{(\ref{#1})}

\begin{document}

\title{Analytical Formulation of the Single-Visit Completeness Joint Probability Density Function}

\author{Daniel Garrett}
\and
\author{Dmitry Savransky\altaffilmark{1}}
\altaffiltext{1}{Carl Sagan Institute, Cornell University, Ithaca, NY 14853}
\affil{Sibley School of Mechanical and Aerospace Engineering, Cornell University, Ithaca, NY 14853}

\begin{abstract}
    We derive an exact formulation of the multivariate integral representing the single-visit obscurational and photometric completeness joint probability density function for arbitrary distributions for planetary parameters. We present a derivation of the region of nonzero values of this function which extends previous work, and discuss time and computational complexity costs and benefits of the method. We present a working implementation, and demonstrate excellent agreement between this approach and Monte Carlo simulation results.
\end{abstract}

\keywords{methods: analytical --- methods: statistical --- planetary systems --- planets and satellites: detection --- techniques: high angular resolution}

\section{Introduction}
Obscurational completeness was introduced by \citet{BrownOC} as a necessary, but not sufficient, condition for detection of an exoplanet. Assuming distributions for semimajor axis and eccentricity of planetary orbits, Brown defined obscurational completeness as the probability of a planet falling outside a telescope's central obscuration, thus becoming potentially observable. This concept was expanded \citep{BrownSVPOC} to include selection effects due to photometric restrictions on exoplanet observability introduced by telescope optics. Completeness can be extended to indirect exoplanet detection methods like reflex astrometry \citep{BrownRA} and account for successive observations  \citep{BrownNew}. Benefits of completeness studies include realistic expectations of mission outcomes based on objective terms for search power and a scientific metric to inform and optimize mission designs \citep{BrownPhO}. These studies have been used in mission analysis and design for a variety of proposed exoplanet observatories \citep{BrownRA, LindlerTPF, SavranskyDRM, SavranskyPFM, SavranskySMD, Stark2014, Brown2015, Stark2015}.

Single-visit completeness is determined by the assumption that an exoplanet is observable if its angular separation from its star is greater than the observatory's inner working angle (IWA) and less than the observatory's outer working angle (OWA) while also being illuminated such that the difference in brightness between the star and planet ($\Delta$mag) is below a limiting value ($\Delta\mathrm{mag}_{0}$). The IWA and OWA represent the minimum and maximum angular separation of the field of view. $\Delta\mathrm{mag}_{0}$ is where unresolvable confusion between the planet signal and noise occur.

For simple cases an analytic functional representation of completeness may be possible. With the exception of \citet{Agol}, previous approaches to calculating completeness have relied on Monte Carlo trials because of the complexity of the assumed distributions. Probability distributions are assumed for the orbital elements and physical properties necessary for determining separation and $\Delta\mathrm{mag}$. A large, equal number of samples (\citet{BrownSVPOC} used 100 million) is generated from each of the distributions. Successive function evaluations are made, including solving Kepler's equation iteratively, leading to the calculation of star-planet separation and $\Delta\mathrm{mag}$ for each set of samples. A two-dimensional histogram of these values is constructed which gives the expectation, or relative frequency of occurrence, of separation and $\Delta\mathrm{mag}$ for each bin of the two-dimensional histogram. Dividing the expectation values by the area of each bin gives the joint probability density function. Integrating with respect to separation and $\Delta\mathrm{mag}$ over this joint probability density function yields a cumulative density function which gives the completeness, or probability that an observatory with given $\Delta\mathrm{mag}_{0}$, IWA, and OWA, observing a specific star for the first time, will detect a planet belonging to the assumed population.

The Monte Carlo trial approach of finding the expectation is analogous to numerical integration. Increasing the number of samples, $n$, in the Monte Carlo trial approach results in reduced error which goes as $O\left(n^{-1/2}\right)$ \citep{Davis}. To reduce the error by one decimal place, the number of samples must be increased by a factor of 100. For one-dimensional numerical integration, the simple Riemann sum, $O\left(n^{-1}\right)$; Newton-Cotes quadrature, better than $O\left(n^{-2}\right)$; or Gaussian quadrature all have better error performance for $n$ sample points than Monte Carlo integration. Multidimensional integrals may be numerically integrated using a composition of one-dimensional integrals or product rules which will also have better error performance than Monte Carlo integration. 

In terms of time complexity, the Monte Carlo trial approach requires sampling of quantities, solving Kepler's equation iteratively, additional function evaluations to get separation and $\Delta\mathrm{mag}$, and sorting these values into a two-dimensional histogram. All of these operations can be performed in polynomial time or better. Numerical integration algorithms require functional evaluations at sample points, determination of weights, multiplying the weights with functional evaluations, and summing these values. All of these operations can be performed in polynomial time. Increasing the dimension of an integral exponentially increases the number of samples required which increases the computational time. Multidimensional integrals higher than about dimension three are better computed using Monte Carlo integration due to the computational time.

The bivariate distribution sampled by Monte Carlo trials is a function of non-independent variables. The completeness distribution must be sampled fully to find any one point of the joint probability distribution accurately. Because of the number of parameters involved and the potential wide range of values these parameters may take, full sampling requires a large number of Monte Carlo trials. Increasing numbers of Monte Carlo trials increases the computational time of accurately determining completeness. For exoplanet mission simulators, it is desirable to produce completeness values quickly without sacrificing accuracy. Numerical integration of lower dimensional integrals would require fewer sample points and give better error and computational time performance for a single point of the joint probability density function.

We present a functional approach to determining single-visit completeness which avoids the undersampling problem inherent in the Monte Carlo simulation. This approach also allows for calculation of a single point of the completeness joint probability density function without simulation of the entire phase space. We begin by presenting the necessary assumptions which allow the description of completeness in functional terms. These functional expressions are made to be as general as possible. We then discuss computational considerations of this approach and provide comparisons to the Monte Carlo trial approach.

\section{Assumptions}\label{sec:assumptions}

\citet{SavranskyPDF} presented a derivation of probability density functions for orbital parameters and observed quantities related to completeness under the following four assumptions:

\begin{enumerate}
    \item Closed Keplerian orbits with negative specific orbital energy approximate planetary orbits
    \item Orbital poles are uniformly distributed on the celestial sphere
    \item Effects due to planet-planet interactions, such as resonant orbits, are ignored
    \item Distance from the observer to the target star is much larger than the distance from the target star to any of its planets
\end{enumerate}
These assumptions allow the description of each orbit with the parameter set $\left( a, e, \psi, \theta, \phi \right)$, where $a$ is the semimajor axis, $e$ is the eccentricity, and $\psi, \theta, \phi$ are Euler angles determining the orientation of the orbit in the observer's reference frame. The true anomaly, $\nu$, gives the position of the planet on its orbit at the time of observation. We will use these same assumptions and their following results to derive the joint probability density function for completeness.

A functional description of the apparent planet-star separation, $s$, and difference in brightness between the star and planet, $\Delta\mathrm{mag}$, are required to derive the completeness joint probability density function. The first and third assumptions result in the familiar Keplerian distance between planet and star,
\begin{equation}
\label{eq:r}
    r = \frac{a\left(1 - e^{2}\right)}{1 + e\cos\nu}.
\end{equation}
The fourth assumption results in an approximation of the star-planet-observer or phase angle $\beta$ giving the apparent planet-star separation as
\begin{equation}
\label{eq:s2}
    s = r\sin\beta.
\end{equation}
\citet{BrownSVPOC} defined the ratio of fluxes between planet and star as
\begin{equation}
\label{eq:FR}
    F_{R} \triangleq \frac{F_{p}}{F_{s}} = p\Phi\left(\beta\right)\left(\frac{R}{r}\right)^2,
\end{equation}
where $p$ and $R$ are the planet's geometric albedo and radius and $\Phi$ is the planet's phase function. The difference in brightness between the star and its planet is given by
\begin{equation}
\label{eq:dmag}
    \Delta\mathrm{mag} = -2.5\log_{10}F_{R}.
\end{equation}
Equations \refeq{eq:s2}, \refeq{eq:FR}, and \refeq{eq:dmag} give the relations between the variables $ \left(p,R,\beta,r\right) $ required for deriving a functional expression for the completeness joint probability density function.

We now turn our attention to assumptions on the distributions of these four quantities. While the quantities $p, R, \beta, $ and $r$ are likely interrelated, we assume that $ p, R, \beta, $ and $ r $ are independent to allow simplifications and the following derivation. \citet{SavranskyPDF} showed that $ \beta $ is sinusoidally distributed regardless of the distribution of any other orbital parameter, dependent only on the second assumption above. All other parameters $\left(a,e,p,R\right)$ are assumed to have probability density functions representative of the planet population of interest and are represented by random variables $\left(\bar{a},\bar{e},\bar{p},\bar{R}\right)$. Under the first three assumptions above and assuming independence between $\bar{a}$ and $\bar{e}$, the probability density function for orbital radius \citep{SavranskyPDF}, $f_{\bar{r}}\left(r\right)$, is given by 
\begin{equation}
\label{eq:fr}
    f_{\bar{r}}\left(r\right) = \frac{1}{\pi}\int_{0}^{\infty}\int_{0}^{1}\frac{r}{a\sqrt{\left(ae\right)^2 - \left(a - r\right)^2}}f_{\bar{e}}\left(e\right)def_{\bar{a}}\left(a\right)da.
\end{equation}
The limits of integration will be from the minimum to the maximum values of semimajor axis and eccentricity of the planet population of interest. Physically realizable solutions occur when the integrand is real, i.e., $\left(ae\right)^2 > \left(a - r\right)^2$.

\section{Derivations}\label{sec:derivations}
\subsection{Completeness Joint Probability Density Function}
The joint probability density function of the variables $ p, R, \beta, $ and $ r $, given the assumptions from Section \ref{sec:assumptions} is
\begin{equation}
\label{eq:fpRbr}
    f_{\bar{p},\bar{R},\bar{\beta},\bar{r}}\left(p,R,\beta,r\right) = f_{\bar{p}}\left(p\right)f_{\bar{R}}\left(R\right)f_{\bar{\beta}}\left(\beta\right)f_{\bar{r}}\left(r\right).
\end{equation}
The completeness joint probability density function, $ f_{\bar{s},\overline{\Delta\mathrm{mag}}}\left(s,\Delta\mathrm{mag}\right) $, will be found in two steps. We will perform a change of variables on Equation \refeq{eq:fpRbr} \citep{Larson} to get a joint probability density function of $ s, \Delta\mathrm{mag}, p, $ and $ R $, $ f_{\bar{s},\overline{\Delta\mathrm{mag}},\bar{p},\bar{R}}\left(s,\Delta\mathrm{mag},p,R\right) $. This distribution will then be marginalized to yield the completeness joint probability density function
\begin{equation}
\label{eq:fsdmag}
    f_{\bar{s},\overline{\Delta\mathrm{mag}}}\left(s,\Delta\mathrm{mag}\right) = \int_{-\infty}^{\infty}\int_{-\infty}^{\infty}f_{\bar{s},\overline{\Delta\mathrm{mag}},\bar{p},\bar{R}}\left(s,\Delta\mathrm{mag},p,r\right)dRdp.
\end{equation}

We begin by defining four new variables as functions of the original variables
\begin{subequations}
\label{eq:g}
    \begin{align}
        s & = g_{1}\left(p,R,\beta,r\right) = r\sin\beta \\
        \Delta\mathrm{mag} & = g_{2}\left(p,R,\beta,r\right) = -2.5\log_{10}\left[p\left(\frac{R}{r}\right)^2\Phi\left(\beta\right)\right] \\
        p & = g_{3}\left(p,R,\beta,r\right) = p \\
        R & = g_{4}\left(p,R,\beta,r\right) = R \,.
    \end{align}
\end{subequations}
All $ g_{i} $ are required to have continuous partial derivatives and $ \Phi\left(\beta\right) $ must be differentiable since the inverse of the Jacobian determinant of Equation \refeq{eq:g} will be used. All $ g_{i} $ satisfy this condition except for $ g_{2} $ when any of the following occur: $ p = 0, R = 0, \beta = 0,\pi, $ and $ r = 0 $. At any of the values violating the condition of continuous partial derivatives, the joint probability density function, $ f_{\bar{s},\overline{\Delta\mathrm{mag}},\bar{p},\bar{R}}\left(s,\Delta\mathrm{mag},p,R\right) $, is set to zero.

The Jacobian determinant of Equation \refeq{eq:g} is given by 
\begin{align}
    J\left(p,R,\beta,r\right) &= \left|
    \begin{array}{cccc}
        \frac{\partial g_{1}}{\partial p} & \frac{\partial g_{1}}{\partial R} & \frac{\partial g_{1}}{\partial \beta} & \frac{\partial g_{1}}{\partial r} \\
        \frac{\partial g_{2}}{\partial p} & \frac{\partial g_{2}}{\partial R} & \frac{\partial g_{2}}{\partial \beta} & \frac{\partial g_{2}}{\partial r} \\
        \frac{\partial g_{3}}{\partial p} & \frac{\partial g_{3}}{\partial R} & \frac{\partial g_{3}}{\partial \beta} & \frac{\partial g_{3}}{\partial r} \\
        \frac{\partial g_{4}}{\partial p} & \frac{\partial g_{4}}{\partial R} & \frac{\partial g_{4}}{\partial \beta} & \frac{\partial g_{4}}{\partial r}
    \end{array} \right| \nonumber \\
    &= \frac{-2.5}{\Phi\left(\beta\right)\ln10}\Phi^\prime\left(\beta\right)\sin\beta - \frac{5}{\ln10}\cos\beta.
\end{align}

$ J\left(p,R,\beta,r\right) $ is required to be non-zero since its inverse will be used. This requirement is violated when $ \beta = \beta^{*} $, where $ \beta^{*} $ maximizes $ \sin^{2}\beta\Phi\left(\beta\right) $ \citep{BrownRV}. As before, when $ \beta = \beta^{*} $, the joint probability density function, $ f_{\bar{s},\overline{\Delta\mathrm{mag}},\bar{p},\bar{R}}\left(s,\Delta\mathrm{mag},p,R\right) $, is set to zero.

We now define the inverse equations to Equation \refeq{eq:g}
\begin{subequations}
    \begin{align}
        p & = h_{1}\left(s,\Delta\mathrm{mag},p,R\right) = p \\
        R & = h_{2}\left(s,\Delta\mathrm{mag},p,R\right) = R \\
        \beta & = h_{3}\left(s,\Delta\mathrm{mag},p,R\right) \\
        r & = h_{4}\left(s,\Delta\mathrm{mag},p,R\right) = \frac{s}{\sin\left(h_{3}\left(s,\Delta\mathrm{mag},p,R\right)\right)}
    \end{align}
\end{subequations}
All of these equations must have a unique solution. $ h_{3} $ and $ h_{4} $ now pose a problem because they may have multiple solutions. Solutions to $ h_{3} $ are the values of $ \beta $ which solve 
\begin{equation}
    \sin^2\beta\Phi\left(\beta\right) = \frac{10^{-0.4\Delta\mathrm{mag}}s^2}{pR^2}.
\end{equation}
If $ h_{3} $ has only one solution, $ \beta = \beta^{*} $ and $ f_{\bar{s},\overline{\Delta\mathrm{mag}},\bar{p},\bar{R}}\left(s,\Delta\mathrm{mag},p,R\right) = 0 $. If $ h_{3} $ has no solution, $ f_{\bar{s},\overline{\Delta\mathrm{mag}},\bar{p},\bar{R}}\left(s,\Delta\mathrm{mag},p,R\right) = 0 $. For the case of multiple solutions we use piecewise definitions for these equations. When two solutions occur, let $ \beta_{1} $ and $ \beta_{2} $ be the two solutions for $ h_{3} $ and $ r_{1} $ and $ r_{2} $ be the corresponding solutions for $ h_{4} $:
\begin{subequations}
    \begin{align}
        \beta_{1} = h_{3}\left(s,\Delta\mathrm{mag},p,R\right), & \;\; 0<\beta_{1}<\beta^{*} \\
        \beta_{2} = h_{3}\left(s,\Delta\mathrm{mag},p,R\right), & \;\; \beta^{*}<\beta_{2}<\pi
    \end{align}
\end{subequations}
and
\begin{subequations}
    \begin{align}
        r_{1} = \begin{cases}
        \frac{s}{\sin\beta_{1}}, & r_{\mathrm{min}} \leq r_{1} \leq r_{\mathrm{max}} \\
        0, & \mathrm{else}
        \end{cases} \\
        r_{2} = \begin{cases}
        \frac{s}{\sin\beta_{2}}, & r_{\mathrm{min}} \leq r_{2} \leq r_{\mathrm{max}} \\
        0, & \mathrm{else}
        \end{cases}
    \end{align}
\end{subequations} where $ r_{\mathrm{min}} $ and $ r_{\mathrm{max}} $ are the minimum and maximum planet distances from the star. The joint distribution function $ f_{\bar{s},\overline{\Delta\mathrm{mag}},\bar{p},\bar{R}}\left(s,\Delta\mathrm{mag},p,R\right) $ is given by the sum of each piece,
\begin{equation}
\begin{split}
    f_{\bar{s},\overline{\Delta\mathrm{mag}},\bar{p},\bar{R}}\left(s,\Delta\mathrm{mag},p,R\right) & = f_{\bar{p},\bar{R},\bar{\beta},\bar{r}}\left(h_{1},h_{2},\beta_{1},r_{1}\right)\left|J\left(h_{1},h_{2},\beta_{1},r_{1}\right)\right|^{-1} \\
    & + f_{\bar{p},\bar{R},\bar{\beta},\bar{r}}\left(h_{1},h_{2},\beta_{2},r_{2}\right)\left|J\left(h_{1},h_{2},\beta_{2},r_{2}\right)\right|^{-1} .
\end{split}
\end{equation}
We now have expressions for all of the functions required in Equation \refeq{eq:fsdmag}.

\subsection{Nonzero Regions of Completeness Joint Probability Density Function}
For a given value of $ s $, $ f_{\bar{s},\overline{\Delta\mathrm{mag}}}\left(s,\Delta\mathrm{mag}\right) $ is only nonzero inside minimum and maximum values of $ \Delta\mathrm{mag} $. We now derive curves for minimum and maximum $ \Delta\mathrm{mag} $ as a function of $ s $ and minimum and maximum values of the assumed planetary population parameters.

To find the minimum $ \Delta\mathrm{mag} $, we insert Equations \refeq{eq:s2} and \refeq{eq:FR} into Equation \refeq{eq:dmag} to get 
\begin{equation}
\label{eq:dmag_s}
    \Delta\mathrm{mag} = -2.5\log_{10}\left[p\Phi\left(\beta\right)\left(\frac{R\sin\beta}{s}\right)^2\right].
\end{equation}
$ \Delta\mathrm{mag} $ is minimized when the expression inside the logarithm in Equation \refeq{eq:dmag_s} is maximized. This leads to the obvious choices of $ p_{\mathrm{max}} $ and $ R_{\mathrm{max}} $, the assumed planetary population parameters limits. To find the value of $ \beta $, we find local extrema which are roots of the following equation \citep{BrownRV}:
\begin{equation}
\label{eq:extrema}
    2\sin\beta\cos\beta\Phi\left(\beta\right) + \sin^{2}\beta\frac{\partial\Phi\left(\beta\right)}{\partial\beta} = 0.
\end{equation}
One root occurs for $ \beta = \beta^{*} $ $ \left(0<\beta^*<\pi\right) $. The other two roots at $\beta = 0$ and $\beta = \pi$ do not occur for all orbital orientations. The extrema for these orbital orientations occur at the planet's closest approach to the star in the plane of the sky, i.e., $ \beta = \sin^{-1}\left(s/r\right) $ or $ \beta = \pi - \sin^{-1}\left(s/r\right) $. The combination of $ \beta^{*} $ and $ s $ will give the minimum value of Equation \refeq{eq:dmag_s} as long as $ r_{\mathrm{min}}\sin\beta^{*} \leq s \leq r_{\mathrm{max}}\sin\beta^{*} $, where $ r_{\mathrm{min}} = a_{\mathrm{min}}\left(1-e_{\mathrm{max}}\right) $ and $ r_{\mathrm{max}} = a_{\mathrm{max}}\left(1+e_{\mathrm{max}}\right) $. Outside of this range, $ \beta $ must correspond to the phase angle at the closest approach. Equation \refeq{eq:dmagmin} summarizes these results.
\begin{equation}
\label{eq:dmagmin}
    \Delta\mathrm{mag}_{min}\left(s\right) = 
    \begin{cases}
        -2.5\log_{10}\left[p_{max}\left(\frac{R_{max}}{r_{min}}\right)^2\Phi\left(\sin^{-1}\left(\frac{s}{r_{min}}\right)\right)\right] & 0 \leq s \leq r_{min}\sin\beta^* \\
        -2.5\log_{10}\left[p_{max}\left(\frac{R_{max}}{s}\right)^2\Phi\left(\beta^*\right)\right] & r_{min}\sin\beta^* \leq s \leq r_{max}\sin\beta^* \\
        -2.5\log_{10}\left[p_{max}\left(\frac{R_{max}}{r_{max}}\right)^2\Phi\left(\sin^{-1}\left(\frac{s}{r_{max}}\right)\right)\right] & r_{max}\sin\beta^* \leq s \leq r_{max}
    \end{cases}
\end{equation}

To find the maximum $ \Delta\mathrm{mag} $, we wish to minimize Equation \refeq{eq:FR}. This leads to the obvious choices of assumed planetary population limits $ p_{\mathrm{min}} $, $ R_{\mathrm{min}} $, and $ r_{\mathrm{max}} $. The value for $ \beta $ should give the smallest value of the phase function $ \Phi\left(\beta\right) $. This occurs for $ \beta = \pi - \sin^{-1}\left(s/r_{\mathrm{max}}\right) $. Equation \refeq{eq:dmagmax} gives this expression.
\begin{equation}
\label{eq:dmagmax}
    \Delta\mathrm{mag}_{max}\left(s\right) = -2.5\log_{10}\left[p_{min}\left(\frac{R_{min}}{r_{max}}\right)^2\Phi\left(\pi -\sin^{-1}\left(\frac{s}{r_{max}}\right)\right)\right]
\end{equation}

\section{Validation of Derived Completeness Joint Probability Density Function}
As a check on the derived completeness joint probability density function, we performed a comparison of Monte Carlo trials to the functional approach derived here. The planetary population variables with their units, minimum and maximum values, and assumed probability density functions are summarized in Table \ref{tab:population}. We assumed log-uniform distributions for semi-major axis, geometric albedo, and planetary radius as zeroth order approximations of the distributions of these quantities based on discovered exoplanets available in catalogs such as \href{http://exoplanet.eu/}{exoplanet.eu} or \href{http://exoplanets.org/}{exoplanets.org}. The log-uniform distribution is given by:
\begin{equation}
\label{eq:loguniform}
    f_{\bar{x}}\left(x\right) = 
    \begin{cases}
        \left(x\ln\left[x_{\mathrm{max}}/x_{\mathrm{min}}\right]\right)^{-1} & x \in \left[x_{\mathrm{min}},x_{\mathrm{max}}\right] \\
        0 & \mathrm{else} \,.
    \end{cases}
\end{equation}
The second assumption in \S\ref{sec:assumptions} requires the mean anomaly to have a uniform distribution. A uniform distribution is given by:
\begin{equation}
\label{eq:uniform}
    f_{\bar{x}}\left(x\right) =
    \begin{cases}
        \left(x_{\mathrm{max}}-x_{\mathrm{min}}\right)^{-1} & x \in \left[x_{\mathrm{min}},x_{\mathrm{max}}\right] \\
        0 & \mathrm{else} \,.
    \end{cases}
\end{equation}
By the second assumption in \S\ref{sec:assumptions}, $ \beta $ is sinusoidally distributed regardless of the distribution of any other orbital parameter \citep{SavranskyPDF}. The sinusoidal distribution is given by:
\begin{equation}
\label{eq:sinusoidal}
    f_{\bar{x}}\left(x\right) = 
    \begin{cases}
        \frac{\sin x}{2} & x \in \left[x_{\mathrm{min}},x_{\mathrm{max}}\right] \\
        0 & \mathrm{else} \,.
    \end{cases}
\end{equation}
We assumed a Rayleigh distribution for eccentricity since it fits Kepler data well \citep{VanEylen2015}. The Rayleigh distribution is given by:
\begin{equation}
\label{eq:Rayleigh}
    f_{\bar{x}}\left(x\right) =
    \begin{cases}
        \frac{x}{c\sigma^{2}}e^{-x^{2}/\left(2\sigma^{2}\right)} & x \in \left[x_{\mathrm{min}},x_{\mathrm{max}}\right] \\
        0 & \mathrm{else}\,.
    \end{cases}
\end{equation}
where $ \sigma $ is a scale parameter and $ c $ normalizes such that the integral from $ x_{\mathrm{min}} $ to $ x_{\mathrm{max}} $ results in 1. We also used the Lambert phase function, \begin{equation}
    \Phi_{L}\left(\beta\right) = \frac{1}{\pi}\left[\sin\beta + \left(\pi-\beta\right)\cos\beta\right].
\end{equation}

\floattable
\begin{deluxetable}{CcCcccc}
\tablecaption{Planetary population distribution \label{tab:population}}
\tablecolumns{6}
\tablenum{1}
\tablewidth{0pt}
\tablehead{
\colhead{Variable} & \colhead{Quantity} & \colhead{Unit} & \colhead{Minimum} & \colhead{Maximum} & \colhead{Distribution}
}
\startdata
a & semi-major axis & AU & 0.5 & 5 & log-uniform \\
e & eccentricity & \cdots & 0 & 0.35 & Rayleigh, $ \sigma = 0.25 $ \\
M & mean anomaly & rad & 0 & $ 2\pi $ & uniform \\
p & geometric albedo & \cdots & 0.2 & 0.3 & log-uniform \\
R & planetary radius & km & 6,000 & 30,000 & log-uniform \\
$ \beta $ & phase angle & rad & 0 & $ \pi $ & sinusoidal \\
\enddata
\end{deluxetable}

\begin{figure}
\figurenum{1}
\label{fig:MC}
    \plotone{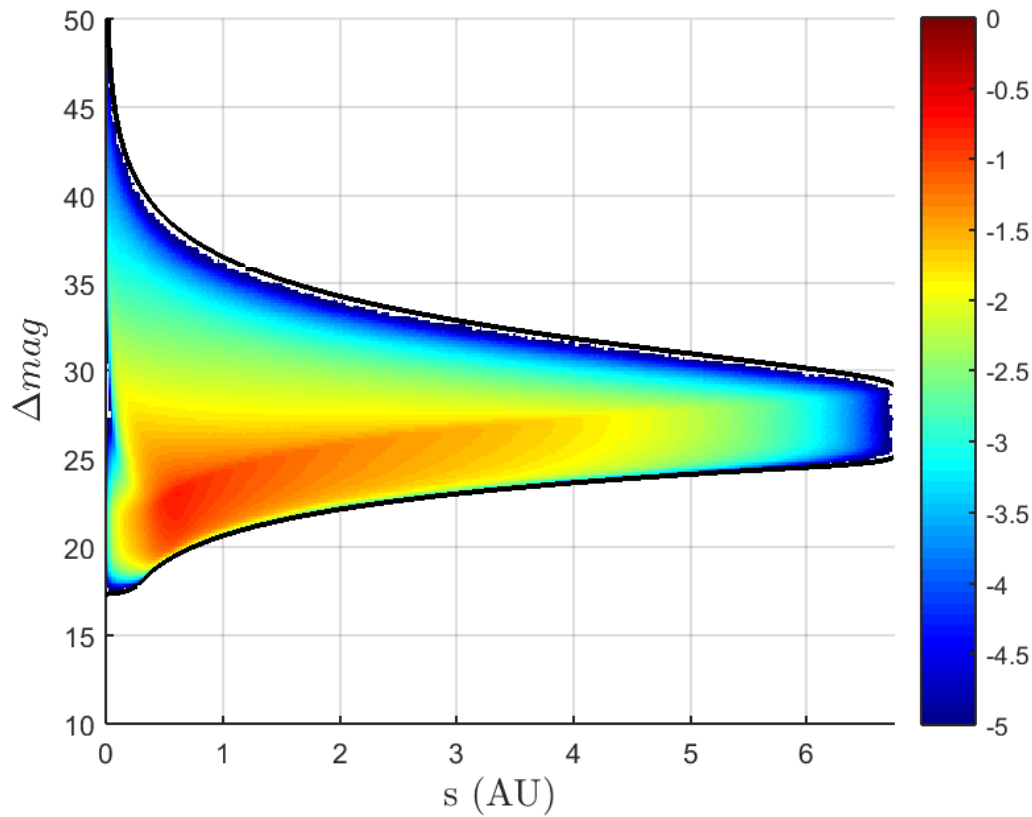}
    \caption{Completeness joint probability density function determined by one billion Monte Carlo trials. The color is log-scaled (base 10) of the probability density $ \mathrm{AU}^{-1}\Delta\mathrm{mag}^{-1} $. The black lines represent minimum and maximum values for which the completeness joint probability density function is nonzero (given by Equations \refeq{eq:dmagmin} and \refeq{eq:dmagmax}). Blank space between these two lines shows that even with one billion trials the Monte Carlo approach does not adequately sample the entire $s$-$\Delta \mathrm{mag}$ phase space.}
\end{figure}

For the Monte Carlo trials, we generated one billion independent identically distributed (IID) samples of the planetary population variables according to their respective probability distributions. We calculated $ s $ and $ \Delta\mathrm{mag} $ for each sample and determined the completeness joint probability density function after sorting the $\left(s,\Delta\mathrm{mag}\right) $ pairs into a $ 400 \times 400 $ grid over the ranges $ 0 \leq s \leq 6.75 $ and $ 10 \leq \Delta\mathrm{mag} \leq 50 $. These computations were performed in parallel on a 4-core 2.3 GHz processor taking $\sim$20 minutes. Figure \ref{fig:MC} shows the resulting completeness joint probability density function with the color log-scaled (base 10). Minimum and maximum $ \Delta \mathrm{mag} $ (Equations \refeq{eq:dmagmin} and \refeq{eq:dmagmax}) are shown. Even with one billion samples, the Monte Carlo trial method does not fill the space between the minimum and maximum $ \Delta \mathrm{mag} $ values. 

\begin{figure}
\figurenum{2}
\label{fig:func}
    \plotone{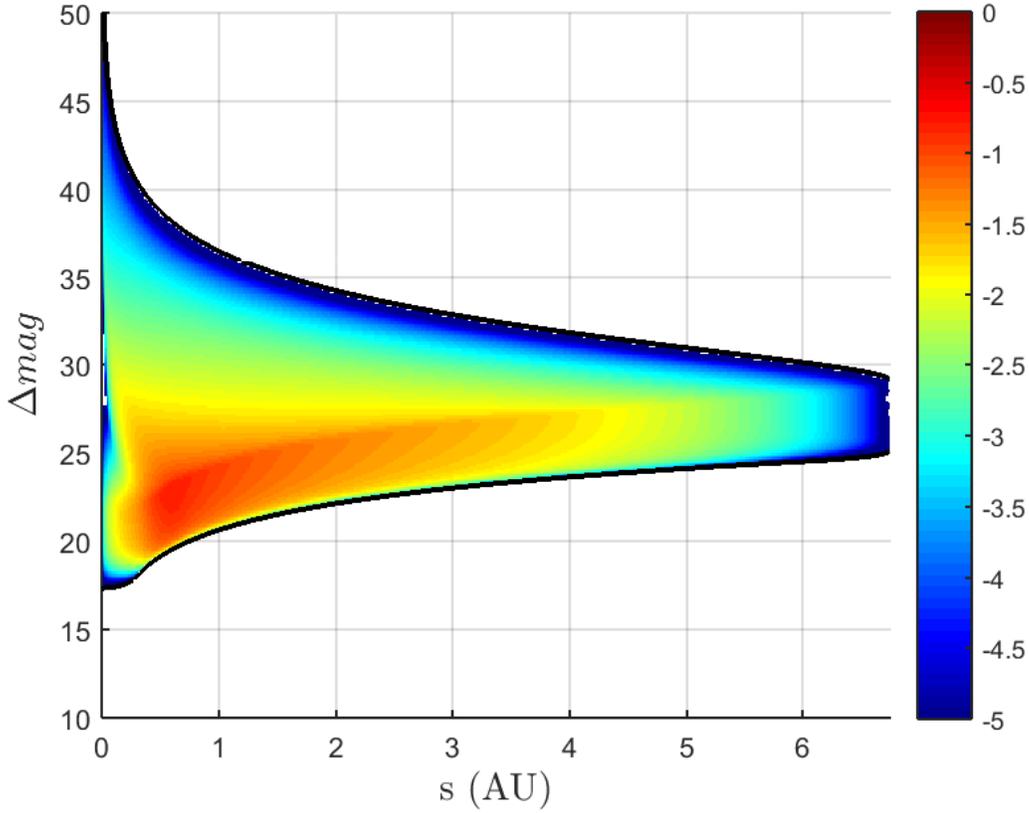}
    \caption{Completeness joint probability density function determined by functions in Section \ref{sec:derivations}. The color is log-scaled in powers of 10 of the probability density $ \mathrm{AU}^{-1}\Delta\mathrm{mag}^{-1} $. The black lines represent minimum and maximum values for which the completeness joint probability density function is nonzero (given by Equations \refeq{eq:dmagmin} and \refeq{eq:dmagmax}). There is no blank space between the lines showing that this approach fills the $s$-$\Delta\mathrm{mag}$ phase space.}
\end{figure}

Using the same $ 400 \times 400 $ grid over $ s $ and $ \Delta\mathrm{mag} $, we performed the calculations outlined in Section \ref{sec:derivations} at each point. These calculations were performed in parallel on the same 4-core 2.3 GHz processor taking $\sim$80 minutes total. Each individual point on the grid took $\sim$0.1 seconds on average to compute. We point out again that completeness requires integration over a region of this joint probability density function. The Monte Carlo approach takes the same amount of computational time whether one point or the entire grid is computed. If higher order numerical integration algorithms are used to compute the double integral giving completeness, the method derived here will give better accuracy and quicker computation times because relatively few functional evaluations are needed compared to fully sampling the space for the Monte Carlo approach. If during the calculation of single visit completeness for a list of 100 target stars by a numerical integration algorithm, the function is evaluated in parallel on the same 4-core 2.3 GHz processor on a $200 \times 50$ grid over $s$ and $\Delta\mathrm{mag}$ (between the minimum possible $\Delta\mathrm{mag}$ and the maximum $\Delta\mathrm{mag}_{0}$ for a population) it will take $\sim$14 minutes. The functional completeness joint probability density function is shown in Figure \ref{fig:func} with the same color scale and minimum and maximum values of $ \Delta \mathrm{mag} $ as in Figure \ref{fig:MC}. Visually, the two Figures look similar. The functional completeness joint probability density function shown in Figure \ref{fig:func}, however, does not have the blank space that the Monte Carlo trial approach in Figure \ref{fig:MC} has. This shows that the functional approach fills the $s$-$\Delta \mathrm{mag}$ phase space completely. 

\begin{figure}
\figurenum{3}
\label{fig:percent}
    \plotone{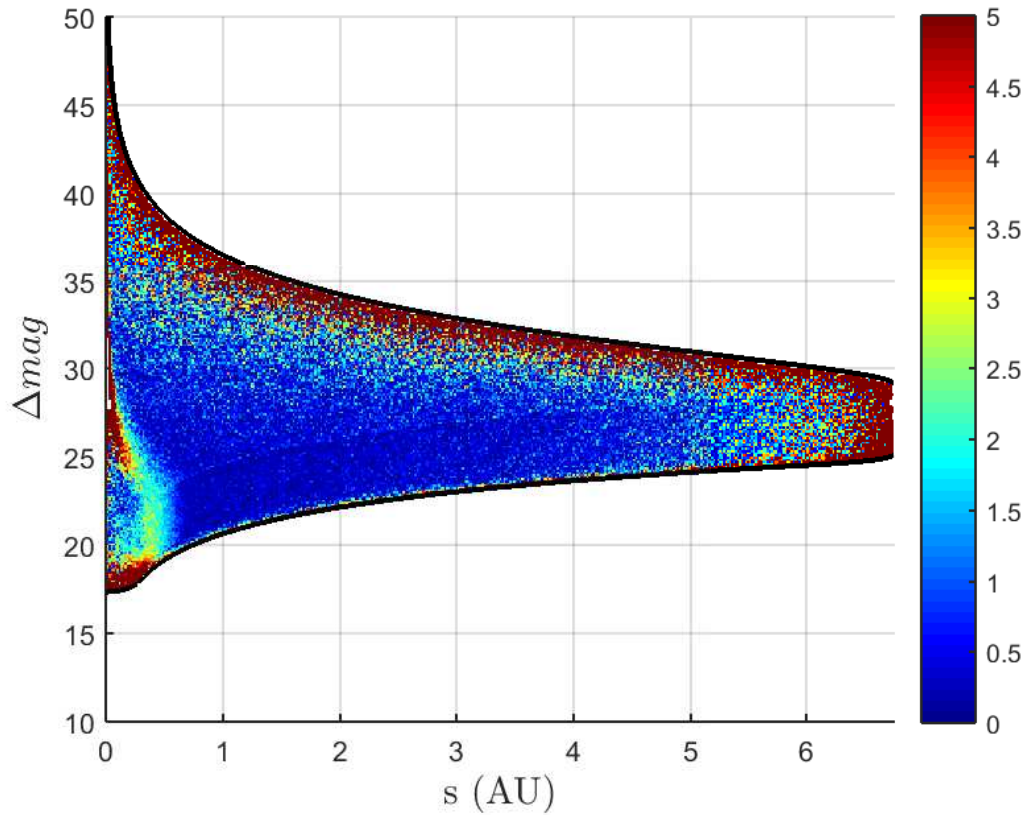}
    \caption{Absolute value of the percent difference from the Monte Carlo approach to the functional approach. The color is in percent. The black lines represent minimum and maximum values for which the completeness joint probability density function is nonzero (given by Equations \refeq{eq:dmagmin} and \refeq{eq:dmagmax}). Agreement between the two approaches is better than 3\% for the majority of the $s$-$\Delta\mathrm{mag}$ plane. Large discrepancies occur where the Monte Carlo approach suffers from undersampling near the minimum and maximum $\Delta\mathrm{mag}$ values.}
\end{figure}

Figure \ref{fig:percent} shows the absolute value of the percent difference of the Monte Carlo trials to the functional approach. We note agreement to better than 3\% for the majority of the nonzero region of the $ s $-$ \Delta\mathrm{mag} $ plane. Larger discrepancies between the two methods occur near the minimum and maximum boundaries where the Monte Carlo approach suffers from undersampling.

\begin{figure}
\figurenum{4}
\label{fig:fs}
    \plotone{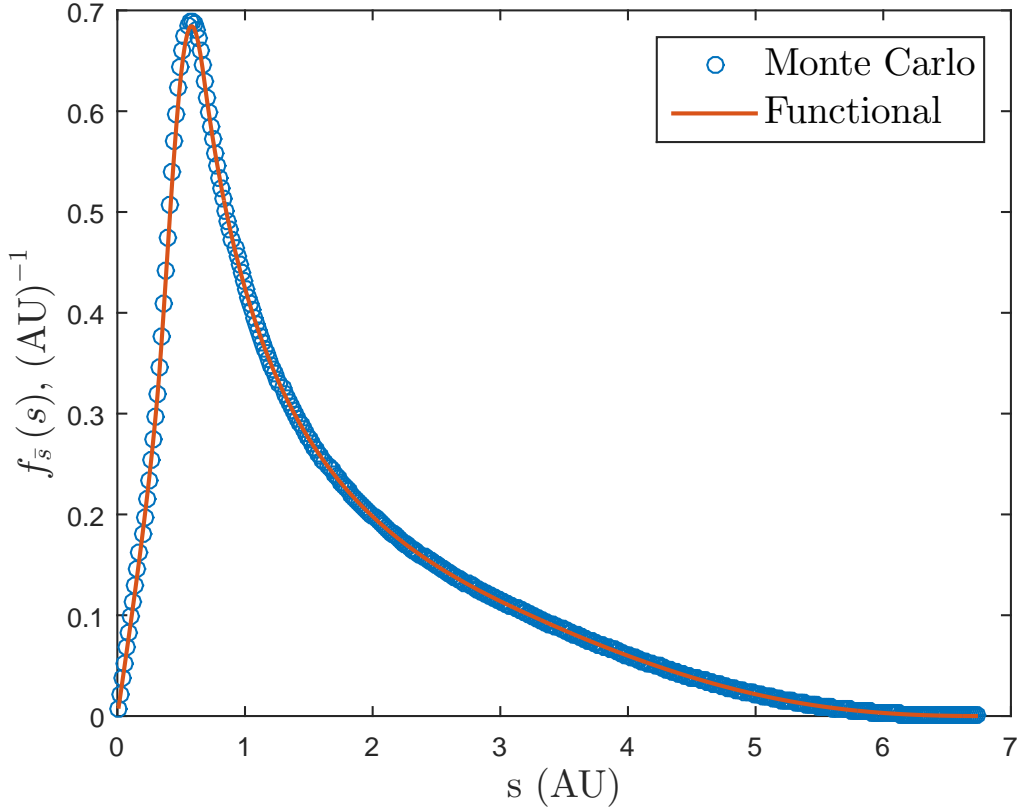}
    \caption{Comparison of Monte Carlo and functional approaches to determine $ f_{\bar{s}}\left(s\right) $.}
\end{figure}

\begin{figure}
\figurenum{5}
\label{fig:fdmag}
    \plotone{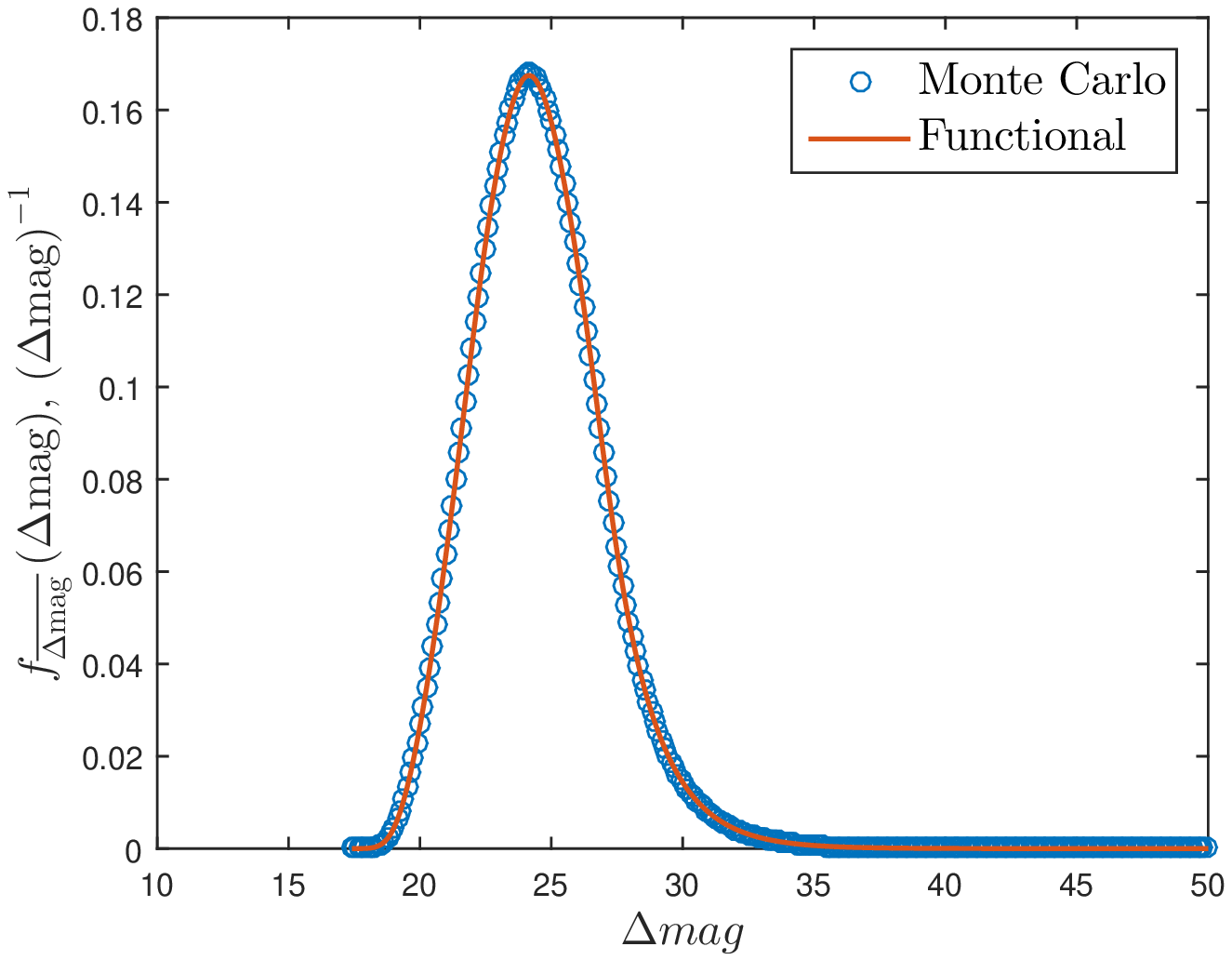}
    \caption{Comparison of Monte Carlo and functional approaches to determine $ f_{\overline{\Delta\mathrm{mag}}}\left(\Delta\mathrm{mag}\right) $.}
\end{figure}

As further validation of the derivation, we performed the same one billion Monte Carlo trials and found probability density functions $ f_{\bar{s}}\left(s\right) $ and $ f_{\overline{\Delta\mathrm{mag}}}\left(\Delta\mathrm{mag}\right) $. The functional approach determines these probability density functions by marginalizing the completeness joint probability density function as in Equations \refeq{eq:fsmarg} and \refeq{eq:fdmagmarg}.
\begin{equation}
\label{eq:fsmarg}
    f_{\bar{s}}\left(s\right) = \int_{-\infty}^{\infty}f_{\bar{s},\overline{\Delta\mathrm{mag}}}\left(s,\Delta\mathrm{mag}\right)d\Delta\mathrm{mag}
\end{equation}
\begin{equation}
\label{eq:fdmagmarg}
    f_{\overline{\Delta\mathrm{mag}}}\left(\Delta\mathrm{mag}\right) = \int_{0}^{\infty}f_{\bar{s},\overline{\Delta\mathrm{mag}}}\left(s,\Delta\mathrm{mag}\right)ds
\end{equation}
Figure \ref{fig:fs} shows the comparison between the probability density function $ f_{\bar{s}}\left(s\right) $ determined by the Monte Carlo approach and the functional approach. Figure \ref{fig:fdmag} shows the comparison between the probability density function $ f_{\overline{\Delta\mathrm{mag}}}\left(\Delta\mathrm{mag}\right) $ determined by the Monte Carlo approach and the functional approach. For each of these probability density functions, the Monte Carlo and functional approaches show very good agreement.

\section{Conclusions}
We have derived an analytical approach for finding the completeness joint probability density function which avoids the undersampling inherent in Monte Carlo approaches and allows computation of a single point of this joint probability density function without simulation of the entire phase space. This allows a quicker, more accurate computation of the double integral giving completeness because the function is evaluated only at points necessary for computing the integral compared to sampling the entire parameter space fully with Monte Carlo trials. This approach is dependent on the assumptions of closed Keplerian orbits; orbital poles distributed uniformly over a spherical volume with respect to the observer; planet-planet interactions neglected; very large distance to target star; and independence of the distributions of geometric albedo, planetary radius, phase angle, and orbital radius. We have shown good agreement between this approach and the Monte Carlo approach. This approach will allow researchers more accurate computation of single-visit photometric and obscurational completeness, which will improve the estimate of the number of extrasolar planets discovered by direct-imaging planet-finding mission simulations and better inform mission design.

\bibliography{funccomp}

\begin{thebibliography}{}
\expandafter\ifx\csname natexlab\endcsname\relax\def\natexlab#1{#1}\fi

\bibitem[{Agol(2007)}]{Agol}
Agol, E. 2007, Monthly Notices of the Royal Astronomical Society, 374, 1271

\bibitem[{Brown(2004{\natexlab{a}})}]{BrownRV}
Brown, R.~A. 2004{\natexlab{a}}, The Astrophysical Journal, 610, 1079

\bibitem[{Brown(2004{\natexlab{b}})}]{BrownOC}
---. 2004{\natexlab{b}}, The Astrophysical Journal, 607, 1003

\bibitem[{Brown(2005)}]{BrownSVPOC}
---. 2005, The Astrophysical Journal, 624, 1010

\bibitem[{Brown(2009{\natexlab{a}})}]{BrownRA}
---. 2009{\natexlab{a}}, The Astrophysical Journal, 699, 711

\bibitem[{Brown(2009{\natexlab{b}})}]{BrownPhO}
---. 2009{\natexlab{b}}, The Astrophysical Journal, 702, 1237

\bibitem[{Brown(2015)}]{Brown2015}
---. 2015, The Astrophysical Journal, 799, 87

\bibitem[{Brown \& Soummer(2010)}]{BrownNew}
Brown, R.~A., \& Soummer, R. 2010, The Astrophysical Journal, 715, 122

\bibitem[{Davis \& Rabinowitz(2007)}]{Davis}
Davis, P.~J., \& Rabinowitz, P. 2007, Methods of numerical integration (Courier
  Corporation)

\bibitem[{Larson \& Shubert(1979)}]{Larson}
Larson, H.~J., \& Shubert, B.~O. 1979, Probabilistic models in engineering
  sciences, Vol.~1 (Wiley)

\bibitem[{Lindler(2007)}]{LindlerTPF}
Lindler, D.~J. 2007, in Optical Engineering+ Applications, International
  Society for Optics and Photonics, 668714--668714

\bibitem[{Savransky(2013)}]{SavranskySMD}
Savransky, D. 2013, in SPIE Optical Engineering+ Applications, International
  Society for Optics and Photonics, 886403--886403

\bibitem[{Savransky {et~al.}(2011)Savransky, Cady, \& Kasdin}]{SavranskyPDF}
Savransky, D., Cady, E., \& Kasdin, N.~J. 2011, The Astrophysical Journal, 728,
  66

\bibitem[{Savransky \& Kasdin(2008)}]{SavranskyDRM}
Savransky, D., \& Kasdin, N.~J. 2008, in SPIE Astronomical Telescopes+
  Instrumentation, International Society for Optics and Photonics,
  70101T--70101T

\bibitem[{Savransky {et~al.}(2010)Savransky, Kasdin, \& Cady}]{SavranskyPFM}
Savransky, D., Kasdin, N.~J., \& Cady, E. 2010, Publications of the
  Astronomical Society of the Pacific, 122, 401

\bibitem[{Stark {et~al.}(2015)Stark, Roberge, Mandell, Clampin,
  Domagal-Goldman, McElwain, \& Stapelfeldt}]{Stark2015}
Stark, C.~C., Roberge, A., Mandell, A., {et~al.} 2015, The Astrophysical
  Journal, 808, 149

\bibitem[{Stark {et~al.}(2014)Stark, Roberge, Mandell, \& Robinson}]{Stark2014}
Stark, C.~C., Roberge, A., Mandell, A., \& Robinson, T.~D. 2014, The
  Astrophysical Journal, 795, 122

\bibitem[{Van~Eylen \& Albrecht(2015)}]{VanEylen2015}
Van~Eylen, V., \& Albrecht, S. 2015, The Astrophysical Journal, 808, 126

\end{thebibliography}
\bibliographystyle{aasjournal.bst}

\end{document}